\documentclass[12pt]{article}

\input{amssymb.sty}

\def\bbc{{\mathbb C}}

\def\bbr{{\mathbb R}}

\def\Tr{\mathop{\rm Tr}\nolimits}
\def\frac#1#2{{{#1}\over{#2}}}
\def\tfrac#1#2{{\textstyle{{#1}\over{#2}}}}

\def\half{\tfrac{1}{2}}

\def\bphi{\mbox{\boldmath $\phi$}}
\def\bomega{\mbox{\boldmath $\omega$}}

\begin{document}

\begin{titlepage}

\baselineskip 24pt

\begin{center}

{\Large {\bf Higgs Fields as Vielbeins of Internal Symmetry Space}}

\vspace{.5cm}

\baselineskip 14pt

{\large CHAN Hong-Mo}\\
h.m.chan\,@\,rl.ac.uk \\
{\it Rutherford Appleton Laboratory,\\
  Chilton, Didcot, Oxon, OX11 0QX, United Kingdom}\\

\vspace{.5cm}

{\large TSOU Sheung Tsun}\\
tsou\,@\,maths.ox.ac.uk\\
{\it Mathematical Institute, University of Oxford,\\
  24-29 St. Giles', Oxford, OX1 3LB, United Kingdom}

\end{center}

\vspace{.3cm}

\baselineskip 14pt

\begin{abstract}

An earlier suggestion that scalar fields in gauge theory may be introduced
as frame vectors or vielbeins in internal symmetry space, and so endowed 
with geometric significance, is here sharpened and refined.  Applied to a 
$u(1) \times su(2)$ theory this gives exactly the Higgs structure of the 
standard electroweak theory. Applied to an $su(3)$ theory, it gives a 
structure  
having much in common with a phenomenological model previously constructed 
to explain fermion mixing and mass hierarchy.  The difference in physical 
outcome for the two theories is here traced to the difference in structure
between the two symmetry groups.

\end{abstract}
\end{titlepage}

\clearpage

\baselineskip 14pt

\setcounter{equation}{0}

\section{Introduction}

Scalar fields as introduced by Higgs \cite{Higgs} and Kibble \cite{Kibble} 
to break gauge symmetries have become an essential feature of almost all
present day particle theories, in particular the current Standard Model
which has great success in correlating existing data.  They are still 
however a bit of an enigma because, apart from the fact that the associated 
bosons have not yet been experimentally observed, they are not known 
theoretically, in stark contrast to the vector gauge fields, to have any 
geometrical significance, which is, to say the least, surprising in a 
theory otherwise so geometrically grounded.  It is therefore of interest 
to explore various theoretical possibilities for assigning to these scalar 
fields a geometrical significance, in the hope of understanding better 
the fundamental structure of gauge theories and gaining some new physical 
insight.   

In an earlier paper while constructing a phenomenological model for fermion
generations \cite{physcons,genmixdsm}, it was suggested that Higgs scalars
may conceivably be given the geometrical significance of frame vectors or 
``vielbeins'' in internal symmetry space, much as the vierbeins familiar in
the theory of relativity, but this suggestion has neither been very precisely 
stated nor yet been vigorously pursued.  It is our purpose in this paper to 
take the proposal further by first clarifying the basic concepts in a more 
general context and then exploring the possible physical consequences.

The considerations so developed are then applied in a companion paper
\cite{prepsm} to the theory with local gauge symmetry $u(1) \times su(2)
\times su(3)$ to deduce many of the properties of the Standard Model  
introduced as empirical inputs in the traditional formulation.

\section{Frame Vectors as Field Variables}

In gravity, one is used to having vierbeins as dynamical variables.  It 
is thus reasonable, requiring no great stretch of imagination, to consider 
also in gauge theories the possibility of introducing frame vectors in 
internal symmetry space as dynamical variables.  Let us first make clear 
what we mean by frame vectors in the gauge theory context, restricting 
ourselves for the moment to $su(N)$ symmetries.\footnote{Here and 
throughout this paper, we denote a gauge symmetry
by its gauge (Lie) algebra whenever there is no necessity to specify 
which among the locally isomorphic Lie groups is to be selected as the 
gauge group, so as to avoid questions of topology inessential for the 
problem in hand.  By convention, lower case letters denote algebras and 
upper case letters groups.}  
A gauge theory is by definition invariant, of course, 
under $x$-dependent changes of the local frames in internal symmetry space.  
A local frame can be specified by giving its orientation with respect to 
some $x$-independent (global) reference frame, via a transformation matrix, 
say $\Omega$, from the global to the local frame, thus:  
\begin{equation}
\Omega = (\omega^{\tilde{a}}_a), \ \ \ \Omega \in SU(N),
\label{Omega}
\end{equation}
with $a = 1,2,..., N$ labelling the rows and $\tilde{a} = 1, 2,..., N$ the 
columns, and it is the columns $\bomega^{\tilde{a}}$ in $\Omega$ here 
we call the frame vectors.  We note that while these (columns of $\Omega$) 
transform under changes of the local frame as fundamental representations 
of $su(N)$, the rows of $\Omega$ on the other hand transform under changes 
of the (global) reference frames as anti-fundamental representations of 
the same algebra.  There are thus two symmetries involved which we shall 
distinguish henceforth as $su(N)$ and $\widetilde{su}(N)$, with the first 
being the local one and the second the global one.

As matters stand, the frame vectors $\bomega^{\tilde{a}}$ are all of 
unit length.  To promote them into fields as suggested, we want to allow 
them the possibility of ranging over all complex values.  This we 
interpret as embedding all of them, or in other words the matrix $\Omega$,
into a linear space and take the coordinates thereof as the actual field 
variables:
\begin{equation}
\Phi = (\phi^{\tilde{a}}_a).
\label{Phi}
\end{equation}
There are of course many ways that one can do so.  The simplest is just 
to take the elements $\omega_a^{\tilde{a}}$ of $\Omega$ themselves, but 
releasing them from all the unitarity constraints that they originally 
satisfy so that each can now range independently over all complex values.  
This means embedding $\Omega$ in $\bbc^{N^2}$ or $\bbr^{2{N^2}}$.

For the case of the symmetry $su(2)$, however, this simple embedding into 
$\bbr^8$ is clearly much too extravagant.  For $su(N)$ symmetries, the
matrix $\Omega$ forms a faithful representation of the group $SU(N)$, so 
that embedding $\Omega$ is the same as embedding the group itself into a 
linear space, and $SU(2)$, being topologically a 3-sphere $S^3$, can be 
embedded already in $\bbr^4$.  In the above language, this means 
releasing $\Omega$ not from all the unitarity constraints it satisfies, but 
only from the condition that, say, its column vectors $\bphi^{\tilde{r}}$
be of unit length while keeping the condition that they should be mutually
orthogonal and of the same length, thus explicitly:
\begin{equation}
\phi_r^{\tilde{2}} = - \epsilon_{rs} (\phi_s^{\tilde{1}})^*.
\label{phi1phi2}
\end{equation}
This still allows each component to range over all complex values, though 
no longer independently, and hence still to qualify as field variables 
as above perceived.  That being the case, we need introduce only one of 
the two vectors $\bphi^{\tilde{1},\tilde{2}}$ as field variable, the other 
will be given by the condition (\ref{phi1phi2}), or in other words that we 
have embedded $\Omega$ into $\bbr^4$ as anticipated, which is in fact 
the minimal embedding for $SU(2)$.

We note that a similar embedding is not available to $SU(3)$, for example.  
One could here of course also reduce the dimension of the embedding space by 
insisting that the vector $\bphi^{\tilde{3}}$, say, be always orthogonal 
to the vectors $\bphi^{\tilde{1}}$ and $\bphi^{\tilde{2}}$:
\begin{equation}
\phi_r^{\tilde{3}} = \epsilon_{rst} (\phi_s^{\tilde{1}})^* 
   (\phi_t^{\tilde{2}})^*,
\label{phi1phi2phi3}
\end{equation}
but this relation is inhomogeneous, implying that $\bphi^{\tilde{3}}$ 
would have a different physical dimension from that of 
$\bphi^{\tilde{1}}$ and 
$\bphi^{\tilde{2}}$, a condition that we cannot physically accept.  The 
best that we can do, it seems, is just to retain from the unitarity 
constraints the condition that the determinant be real.  When cast in 
the form:
\begin{equation}
\det \Phi = (\det \Phi)^*,
\label{detreal}
\end{equation}
the condition is homogeneous, thus not suffering from the same objection
against (\ref{phi1phi2phi3}) above, and multilinear, thus allowing the
independent variables to achieve all complex values as demanded.  While
retaining this among all the unitarity constraints on $\Omega$ reduces 
the dimension only by 1 from that of the simple embedding, it seems to 
be for $su(3)$ already the smallest that is physically admissible. 
Similar considerations apply to all $su(N),\ N \geq 3$. 

The smaller the embedding means the smaller the number of scalar fields that
have to be introduced.  For reasons of economy, therefore, it seems to us
reasonable to choose, if possible, the smallest embedding when promoting
frame vectors into fields, and we shall do so.  The difference in the minimal
embedding noted above between $SU(2)$ and $SU(3)$ will then, as we shall
see, play a significant role in the present framework in giving the very 
different physics emerging from the electroweak theory on the one hand 
and from QCD on the other.

The field variables $\bphi^{\tilde{a}}$ so introduced, whether obtained 
in the general $SU(N)$ or in the special $SU(2)$ case from ``minimal'' 
embedding, transform in the same way under gauge transformations as the 
frame vectors $\bomega^{\tilde{a}}$, i.e.\ as the fundamental representation 
of the gauge symmetry, but under Lorentz transformations they are space-time 
scalars.  In other words, by promoting the frame vectors in the internal 
symmetry space into dynamical variables (which we shall refer to henceforth 
as ``framons'' to avoid confusion) we have already smuggled into the gauge 
framework scalar fields with some properties of the Higgs fields we require.

\section{The Invariant Action}

We next proceed to construct an action for the framon fields so introduced.
Apart from Lorentz invariance, we must require of course that the action 
be invariant under local gauge transformations of $su(N)$.  Moreover, 
since physics should presumably also not be affected by the choice of the 
reference frame, it would seem that we should require the action to be also
invariant under global transformations in $\widetilde{su}(N)$.  Together, 
these invariance properties impose stringent restrictions on the actions 
one can construct.  Indeed, one seems to be limited then only to actions of 
the following form:
\begin{equation}
{\cal A} = - \frac{1}{4} \int \Tr\,(F_{\mu\nu} F^{\mu\nu})
          + \int \Tr\,((D_\mu \Phi)^\dagger (D^\mu \Phi))
          + \int V[\Phi].
\label{calA}
\end{equation}
The first term is the usual gauge field action which is of course invariant
under local gauge transformations in $su(N)$, and also invariant trivially
under global transformations in $\widetilde{su}(N)$ since on these it does 
not depend.  The second term, which is the 
kinetic energy term of the $\Phi$ field, is invariant 
under local gauge transformation because of the covariant derivative $D_\mu$ 
defined as usual as:
\begin{equation}
D_\mu = \partial_\mu - ig A_\mu,
\label{Dmu}
\end{equation}
with $A_\mu$ being the gauge potential corresponding to $F_{\mu\nu}$, but
also under global transformation under $\widetilde{su}(N)$ because of the 
trace which is actually taken over $\tilde{a}$ indices.  The third term is
the potential $V[\Phi]$ representing the self-interaction of the $\Phi$
scalar fields, which, because of the required invariance, is limited to
the following form up to fourth order (for renormalizability):
\begin{equation}
V[\Phi] = - \mu\,\Tr\,(\Phi^\dagger \Phi) + 
\lambda\, \left(\Tr\,(\Phi^\dagger \Phi) \right)^2
          + \kappa\, \Tr\,(\Phi^\dagger \Phi \Phi^\dagger \Phi)\,,
\label{VPhi}
\end{equation}
and depends on only three real parameters, $\mu, \lambda, \kappa$.  The 
signs of these parameters have no meaning at present but are labelled such 
only in anticipation of future application.  
  
For the special case of $su(2)$, if we choose the minimal embedding,
i.e.\ 
insist on the orthogonality condition (\ref{phi1phi2}) being satisfied, 
one of the two doublets $\bphi^{\tilde{1},\tilde{2}}$ can be eliminated,
leaving only say $\bphi^{\tilde{1}} = \bphi$ as variable.  A 
straightforward calculation then shows that the action in (\ref{calA}) 
reduces to just:
\begin{equation}
{\cal A} = - \frac{1}{4} \int \Tr\,(F_{\mu\nu} F^{\mu\nu})
           + 2 \int (D_\mu \bphi)^\dagger (D^\mu \bphi) + \int V^{MH}[\bphi]
\label{calASU2}
\end{equation}
where we notice that the framon potential $V[\Phi]$ reduces to  the 
familiar Mexican hat potential:
\begin{equation}
 V^{MH}[\bphi] = - \mu'|\bphi|^2 + \lambda'(|\bphi|^2)^2,
\label{VMHPhi}
\end{equation}
with $\mu' = 2 \mu,\ \lambda' = 4 \lambda + 2 \kappa$.  Alternatively, there 
being really only one independent doublet of framon fields, one can choose 
to start with only say $\bphi^{\tilde{1}} = \bphi$ as variable to 
construct an action invariant under local $su(2)$ transformations getting
directly (\ref{calASU2}) as the result.  The potential $V^{MH}[\Phi]$
depends only on the length of $\bphi$ (a column vector of $\Phi$), 
which however, by (\ref{phi1phi2}), is the same as the length of the rows 
of $\Phi$ and is therefore automatically invariant under $\widetilde{su}(2)$
also.  Or, equivalently, $|\bphi|^2 = ({\rm Re}(\phi_1))^2 + 
({\rm Im}(\phi_1))^2
+ ({\rm Re}(\phi_2))^2 + ({\rm Im}(\phi_2))^2$ is seen to be invariant 
under $so(4)$ which 
is the same as $su(2) \times \widetilde{su}(2)$.

\section{Confinement Picture of Symmetry Breaking}

The action (\ref{calASU2}) with the parameters $\mu', \lambda'$ in $V^{MH}$
both positive is often given as an example of a theory with a local $su(2)$ 
symmetry which is spontaneously broken.  However, as first pointed out by 
't~Hooft \cite{tH} and by Banks and Rabinovici \cite{BR}, and re-emphasized
recently by 't~Hooft \cite{tH1}, it can equally be interpreted as a theory 
in which the local $su(2)$ symmetry confines and remains exact. In this 
confinement picture, what appear as particles observable by
experiment, such 
as the Higgs boson and the massive vector bosons (and eventually also the 
leptons and quarks when fermions are introduced), are all $su(2)$ singlets, 
being compound states, bound or confined by the $su(2)$ symmetry, of the 
fundamental scalar, i.e.\ ``framon'', fields $\phi_r^{\tilde{r}}$ with their 
own conjugates, with the gauge fields (or with some fundamental fermion fields 
yet to be introduced). So long as one assumes, however, as the cited authors 
did, that the confinement by $su(2)$ is very deep, much deeper, say, than the 
confinement by colour, so as to be inaccessible by present experiment, then 
these particles will appear as elementary, giving in present usage the same 
result in practical terms as the conventional picture of symmetry breaking. 
Some may find, as we do, the confinement picture physically more appealing,
which is in any case more convenient to use in certain circumstances than 
the conventional interpretation.

As originally formulated, the confinement picture does not depend of course
on the framon idea suggested here.  However, to realise the picture starting 
from (\ref{calASU2}), one relies  first on the introduction via 
(\ref{phi1phi2}) of the subsidiary fields $\phi_r^{\tilde{2}}$ with the 
original fields $\bphi$ in (\ref{calASU2}) taken as $\phi_r^{\tilde{1}}$, 
and secondly on the existence of an $\widetilde{su}(2)$ symmetry in the 
Mexican hat potential as explained 
at the end of the last section.  Neither of 
these concepts were there originally in the conventional construction of 
(\ref{calASU2}) as an action locally invariant under $su(2)$ of the single 
scalar field $\bphi$, and may thus appear as somewhat incidental.  On 
the other hand, starting with the framon idea proposed above, both the 
$\widetilde{su}(2)$ symmetry and the fields $\phi_r^{\tilde{r}}$ were there 
at the onset and would thus appear, together with the confinement picture,
to be more natural.  Moreover, they have both their parallels in 
other $su(N)$ theories, which then suggest how the confinement picture can 
be generalized to these theories, a subject that will be taken up again 
later.  For future reference, therefore, we shall recast in the rest of 
this section the arguments of 't~Hooft
\cite{tH,tH1} and Banks and Rabinovici 
\cite{BR} for the $su(2)$ theory in the language of the framon proposal.

Let us start then with the invariant action for framons of (\ref{calA}),
specialized in the case of present interest to $su(2)$.  This being by 
construction invariant under local $su(2)$, we can choose to work in the
gauge where $\bphi^{\tilde{1}}$ points in the up direction and is real,
the 3 degrees of freedom in $su(2)$ being just sufficient for this to be
done.  The minimal embedding condition (\ref{phi1phi2}) then implies that
$\bphi^{\tilde{2}}$ would point in the down direction, is real also and
has the same length.  In other words, we have, by a gauge transformation
$\Omega_0$:
\begin{equation}
\Phi = \Omega_0 \Phi_0 
    = \Omega_0 \left( \begin{array}{cc} \rho & 0 \\ 0 & \rho \end{array}
      \right),
\label{Phi0}
\end{equation}
with $\rho$ real and $\Phi_0$ diagonal.  Recalling now the geometric
meaning of $\Phi$ as the transformation matrix from the local to the
global reference frame, we see that for $\Omega_0$ as for $\Phi$, the
rows refer to the local frame but the columns to the global reference
frame.  It is the transformation which rotates the local frame vectors
so as to be aligned to the global reference frame vectors.  We can call 
this the ``locked gauge'' in which the local and global frames are locked 
together in direction.  

Under  the gauge transformation $\Omega_0$, 
we have as usual:
\begin{eqnarray}
A_\mu & \longrightarrow & \tilde{A}_\mu = \frac{i}{g}\, \Omega_0^{\dagger}
   D_\mu \Omega_0, \label{amulongarrow}\\
D_\mu \Phi & = &  \Omega_0 (\partial_\mu - i g
   \tilde{A}_\mu) \Phi_0.
\label{Atilde}
\end{eqnarray}
The action (\ref{calA}) being invariant by construction will remain of
the same form with just $\Phi$ replaced by $\Phi_0$ and $A_\mu$ replaced
by $\tilde{A}_\mu$.  But this is exactly the same as the action one would
obtain in the symmetry breaking picture by choosing the vacuum to be such
as to have $\bphi^{\tilde{1}}$ pointing in the up direction and be real,
except for the replacement of $A_\mu$  by $\tilde{A}_\mu$.

Alternatively one can, so as to be on more familiar grounds, eliminate 
$\bphi^{\tilde{2}}$ via (\ref{phi1phi2}) from consideration, reducing the 
action from (\ref{calA}) to (\ref{calASU2}).  The gauge transformation 
(\ref{amulongarrow}) on the reduced action gives then the standard $su(2)$ 
action in the symmetry breaking picture when the vacuum is chosen to have 
the one remaining scalar doublet real and pointing in the up direction, 
but again with $A_\mu$ replaced by $\tilde{A}_\mu$.

In either case, since the actions are formally the same, one would obtain
the same result, say in perturbation expansion, 
provided that $\tilde{A}_\mu$ can be interpreted
as a point particle.  What does this $\tilde{A}_\mu$ field represent?  We
note from (\ref{amulongarrow}), 
recalling that $\Omega_0$ carries $su(2)$ indices 
only on its left, that $\tilde{A}_\mu$ has no unsaturated $su(2)$ indices,
i.e.\ it is an $su(2)$ singlet.  
To see what it represents, let us expand $\rho$ about its vacuum value, thus
$\rho = F + h_1$, with  $F = \sqrt{\mu'/2 \lambda'}$ being the minimum of
the Mexican hat potential, obtaining for $\Phi_0$:
\begin{equation}
\Phi_0 = \left( \begin{array}{cc} F + h_1 & 0 \\
                                  0 & F + h_1 \end{array} \right),
\label{Phi0exp}
\end{equation}
and hence
\begin{equation}
\Phi^{\dagger} D_\mu \Phi = - ig F^2 \tilde{A}_\mu + O(h_1/F).
\label{PhiDPhiexp}
\end{equation}
One sees then that $\tilde{A}_\mu$ can be interpreted to leading order
in the expansion as a $p$-wave bound state of $\Phi$ with it conjugate
$\Phi^{\dagger}$ obtained by $su(2)$ confinement. Similarly, from:
\begin{equation}
\Phi^{\dagger} \Phi = F^2 + 2 F h_1 + O(h_1/F).
\label{Higgs}
\end{equation}
one sees that the Higgs boson $h_1$ is to be interpreted as an $s$-wave
bound state of $\Phi$ with $\Phi^{\dagger}$.  Thus one has obtained the
same results in the confinement picture as in the symmetry breaking picture, 
only with a different interpretation for the Higgs and vector particles 
as bound states in $su(2)$ confinement.

\section{The Electroweak Theory}

We now turn to an actual physical situation and consider the electroweak
theory.  Let us pretend for the moment that one has never heard before of
the theory and ask, when faced with a theory with gauge symmetry $u(1) 
\times su(2)$, how one would implement the idea of having frame vectors 
introduced as fields, as was done before for $su(N)$.  

Suppose then we are given a gauge theory with gauge potentials $a_\mu$ in 
$u(1)$ and $A_\mu = A_\mu^i \tau_i/2$ in $su(2)$, and the standard field 
action:
\begin{equation}
{\cal A}_F = - \frac{1}{4} \int d^4 x f_{\mu\nu} f^{\mu\nu}
             - \frac{1}{4} \int d^4 x \Tr\,(F_{\mu\nu} F^{\mu\nu})
\label{calAFEW}
\end{equation}
with
\begin{equation}
f_{\mu\nu} = \partial_\nu a_\mu - \partial_\mu a_\nu,
\label{fmunuab}
\end{equation}
and 
\begin{equation}
F_{\mu\nu} = \partial_\nu A_\mu - \partial_\mu A_\nu +i g_2 [A_\mu, A_\nu],
\label{Fmunu}
\end{equation}
and that we are required, by our proposed criterion, to introduce as field 
variables, in addition to $a_\mu$ and $A_\mu$, also the frame vectors in 
the internal symmetry space.  What should these be in the present case when 
the symmetry is not simple as in the $su(N)$ cases previously considered, 
but is instead a product symmetry $u(1) \times su(2)$?

Let us first see what would be the analogue of frame vectors in a $u(1)$
theory.  Frame vectors are there to specify the orientation of the local 
frame relative to some global reference frame.  For the $u(1)$ theory,
orientation means just a phase, and relative orientation just the phase 
difference, thus:
\begin{equation}
\omega = e^{i g_1 (\alpha - \tilde{\alpha})},
\label{phiu1}
\end{equation} 
where $\alpha$ can depend on $x$ but $\tilde{\alpha}$ is $x$-independent.
We have here taken for granted that the gauge group is compact, i.e.\ a
$U(1)$ with ``size'' $2 \pi/g_1$.  Under a local change of frame, the 
frame vector $\omega$ transforms by a $u(1)$ transformation effected by
multiplication by say $\exp i g_1 \delta \alpha$, and, under a global
change of reference frame, by a $\tilde{u}(1)$ transformation effected
by multiplication by $\exp -i g_1 \delta \tilde{\alpha}$.

That being the case, what would a frame vector be for the product symmetry
$u(1) \times su(2)$?  It would have to be a representation of the algebra
$u(1) \times su(2)$ to be constructed out of the representations of $u(1)$
and $su(2)$ for respectively the frame vectors of the component symmetries.
One can in principle take either the sum or the product representation.  
We propose to 
take the product, this being the smaller of the two: $1 + 2
> 1 \times 2$.  The smaller the representation for the local group means
the smaller the
number of $x$-dependent scalar fields that one has eventually
to introduce for framons.  Thus choosing the smaller representation here
is the same in purpose as insisting on the minimal embedding above to 
economize on the number of scalar fields.  With the choice of the product
representation, a frame vector of the $u(1) \times su(2)$ theory will then
appear as a 2-vector:
\begin{equation}
\bomega = \left( \begin{array}{c} \omega_1 \\ \omega_2 \end{array} \right)
\label{omegau12}
\end{equation}
which transforms as a doublet under local $su(2)$ and carries at the same 
time a $u(1)$ charge.

What $u(1)$ charge should $\bomega$ carry?  The analogue in $u(1)$ for 
the doublet in $su(2)$ as the fundamental representation we interpret as
the representation with the minimal quantized charge. The value of the
minimal charge depends on what is taken as the gauge group of the theory.
So far only the gauge algebra is specified as $u(1) \times su(2)$, but
several groups share the same algebra: $U(1) \times SU(2),\ U(1) \times
SO(3),\ U(2)$, where $U(1) \times SO(3)$ has no doublet representations
and can be discarded right away for our 
present consideration.  Of the remaining 2
groups, $U(1) \times SU(2)$ double covers $U(2)$ so that the latter can
be considered the smaller.  Let us choose then the latter, true to the 
spirit of economy maintained throughout, although in this case the
choice of gauge group, in contrast to the choice of representations and 
of embeddings above, does not affect the number of scalar fields that
have to be introduced.  With $U(2)$ as gauge group, the minimal quantized
$u(1)$ charge is $g_1/2$ \cite{ourbook}, and one obtains for the frame
vector $\bomega$ the $u(1)$ charge $\pm g_1/2$, where the ambiguity in 
sign is due to the fundamental representation $\bf 2$ and the antifundamental 
representation $\bf \bar{2}$ of $su(2)$ being equivalent.

The actual phase of the frame vector $\bomega$ depends also on the global
reference phase, i.e.\ the reference frame for the $u(1)$ symmetry, and
will change under a change of this phase.  In other words, $\bomega$
would be assigned a $\tilde{u}(1)$ charge also, which as explained in an
earlier paragraph would have a value opposite to that of the $u(1)$
charge.  Thus under a $u(1)$ transformation represented by the phase
change $\exp i g_1 \delta \alpha$ 
or a $\tilde{u}(1)$ transformation represented by
the phase change $\exp - i g_1 \delta \tilde{\alpha}$, 
the frame vector $\bomega$ or equivalently
the framon field $\bphi$ would transform as:
\begin{equation}
\bphi \stackrel{u(1)}{\longrightarrow} e^{\pm i g_1 \delta \alpha/2} 
\bphi;  \ \ \ 
\bphi \stackrel{\tilde{u}(1)}{\longrightarrow} 
e^{\mp i g_1 \delta \tilde{\alpha}/2} \bphi.
\label{bphiu1}
\end{equation}

We need 2 such framons, say $\bphi^{\tilde{1}}$ and $\bphi^{\tilde{2}}$ 
which, by minimal embedding, are to be related by (\ref{phi1phi2}).  If
given definite $u(1)$ and $\tilde{u}(1)$ charges as in (\ref{bphiu1}),
then their charges will have to be opposite.  They can be taken as:
\begin{eqnarray}
\bphi^{\tilde{1}} & = & \bphi^{(-)}; \ \ u(1)\ {\rm charge} = - \half g_1;
   \ \ \tilde{u}(1)\ {\rm charge} = \half g_1, \nonumber \\
\bphi^{\tilde{2}} & = & \bphi^{(+)}; \ \ u(1)\ {\rm charge} = \half g_1;
   \ \ \tilde{u}(1)\ {\rm charge} = - \half g_1, 
\label{2framons}
\end{eqnarray}
one of which can be eliminated by (\ref{phi1phi2}) in terms of the other
as independent variable.  This is not the most general choice, however, 
for neither basis framon need be an eigenstate of the $u(1)$ or the 
$\tilde{u}(1)$ charge.  One can choose instead:
\begin{eqnarray}
\bphi^{\tilde{1}} & = & \alpha^{(-)} \bphi^{(-)} + \alpha^{(+)} \bphi^{(+)} \\
\bphi^{\tilde{2}} & = & - (\alpha^{(+)})^* \bphi^{(-)} + (\alpha^{(-)})^* 
   \bphi^{(+)},
\label{su2tphi}
\end{eqnarray}
with
\begin{equation}
|\alpha^{(-)}|^2 + |\alpha^{(+)}|^2 = 1.
\label{alpha+-}
\end{equation}
i.e.\ effect an $\widetilde{su}(2)$ transformation on (\ref{2framons}), and
they would still satisfy the minimal embedding condition (\ref{phi1phi2}).
However, the theory being supposed to be $\widetilde{su}(2)$ invariant,
one can always choose to work in the $\widetilde{su}(2)$ gauge (global)
where the basis framons are as in (\ref{2framons}) above.  Indeed, since 
the theory is also $su(2)$ invariant, one can also choose to work in the
gauge when the local and global frames are aligned, in which case the
framon matrix will be diagonal and symmetric between the dual (i.e.\  
with and without tilde)
symmetries, with the charge assignments:
\begin{equation}
\Phi = \left( \begin{array}{cc} [- \half, \half] & 0 \\ 0 & [\half, 
- \half]
   \end{array} \right),
\label{Phiucharge}
\end{equation}
where the first number inside the square brackets denotes the $u(1)$
charge, and the second the $\tilde{u}(1)$ charge.  However, to keep
the invariance under $\widetilde{su}(2)$ explicit, it is often
convenient to work with a general choice of $\bphi^{\tilde{1}}$ and 
$\bphi^{\tilde{2}}$ which need not be eigenstates of the $u(1)$ or the 
$\tilde{u}(1)$ charge.

Having specified the framon fields to be introduced into the theory, our
next task is to construct an action which is now required to be invariant
locally under $u(1) \times su(2)$ and globally under $\tilde{u}(1) \times 
\widetilde{su}(2)$.  

The gauge field action (\ref{calAFEW}) we started with is by construction 
already invariant under local $u(1) \times su(2)$ transformations.  It is 
also invariant trivially under global $\tilde{u}(1) \times \widetilde{su}(2)$ 
transformations because on these it simply does not depend.  It is thus 
already acceptable for our present theory.  

Next, the potential (\ref{VPhi}) was constructed to be invariant under 
$su(2)\times \widetilde{su}(2)$.  It is invariant also under $u(1)$
and $\tilde{u}(1)$ since the phases will just cancel.  So we have in 
(\ref{VPhi}), or equivalently, when $\bphi^{\tilde{2}}$ is eliminated by the
condition (\ref{phi1phi2}) in favour of $\bphi^{\tilde{1}} = \bphi$, 
the Mexican
hat potential (\ref{VMHPhi}), already the potential with the invariance 
required.

This leaves only the kinetic energy term for the framon field, for
which we notice that the corresponding second term in (\ref{calA})
would already suffice provided we replace the covariant derivative in
(\ref{Dmu}) by:
\begin{equation}
D_\mu = \partial_\mu - ig q\, a_\mu - ig_2 A_\mu,
\label{qcoder}
\end{equation}
with $q$ the $u(1)$ charge operator, the resulting expression being
then invariant under both local $u(1) \times su(2)$ and global 
$\tilde{u}(1) \times \widetilde{su}(2)$.  In particular, in terms of
the framons with definite $u(1)$ charges, 
\begin{equation}
{\cal A}_{KE} = \int d^4 x\, 
\left((D_\mu \bphi^{(+)})^{\dagger} (D^\mu \bphi^{(+)}) +
(D_\mu \bphi^{(-)})^{\dagger} (D^\mu \bphi^{(-)}) \right)\,,
\label{calAKEEW}
\end{equation}
with
\begin{equation}
D_\mu \bphi^{(\pm)} = (\partial_\mu \mp i \frac{g_1}{2} a_\mu
                           - i g_2 A_\mu) \bphi^{(\pm)}.
\label{DmuEW}
\end{equation}
Further, a direct calculation in the gauge where $\bphi^{(\pm)}$ are
real and point in the up (down) direction readily shows that, because
of (\ref{phi1phi2}), the two terms in (\ref{calAKEEW}) are in fact
identical and add up to just:
\begin{equation}
{\cal A}_{KE} =2 \int d^4 x 
(D_\mu \mbox{\boldmath $\phi^{(+)}$})^{\dagger} (D^\mu \mbox {\boldmath  
   $\phi^{(+)}$}).
\label{calAKEEWC}
\end{equation}

Together then for the bosonic sector, we have for the action of our $u(1)
\times su(2)$ theory:
\begin{equation}
{\cal A} = {\cal A}_F + {\cal A}_{KE} + \int d^4x\, V^{MH}[\Phi],
\label{calAEW}
\end{equation} 
which is the same as the standard electroweak action in the conventional 
formulation.  It thus seems not only that the Higgs field of the electroweak 
theory can be interpreted as a frame vector in internal symmetry space as 
suggested, and hence be given a geometrical significance that it previously 
lacked, but that, starting with the assumption that frame vectors are to be 
introduced as dynamical variables as part of the gauge framework, one is led, 
modulo arguments on minimality, uniquely to the standard electroweak 
action as the result.

The claim for uniqueness, however, should not be given undue significance. 
The framon idea as conceived at present is insufficient by itself to specify
uniquely which scalar fields are to be introduced for product symmetries like
$u(1) \times su(2)$.  What one did then was to invoke what one might call
the ``principle of minimality'' to resolve the ambiguities in the small
number of choices available.  This ``principle'' was invoked 3 times: first
in the choice of embedding of the frame vectors in $\bbr^N$ to promote 
them into fields (in common with the case of the simple $su(2)$ theory), 
secondly in the choice of the product over the sum representation for 
$u(1) \times su(2)$, and thirdly in the choice of $U(2)$ 
as the gauge group.  The first 2 economise on the number of scalar fields 
that have to be introduced, while the third economises on the number of 
admissible representations.  But one has no good physical reason to give 
for why nature should opt for these economies.  Besides, since one already 
knows the electroweak theory and hence the answer one wants, one could 
probably replace this ``principle of minimality'', if it did not work,
by some other equally plausible criterion.  Nevertheless, we find it 
interesting that an insistence on economy, which is undoubtedly a good 
thing, does lead us uniquely to the correct answer. 

The action (\ref{calAEW}) with $\mu, \lambda > 0$ in $V^{HM}$ is usually
interpreted as representing a theory with its local $su(2)$ gauge symmetry
spontaneously broken.  But, as in the example of the preceding section, it 
can equivalently be interpreted as a theory in which the local $su(2)$ 
symmetry confines.  This follows, as in the previous example, by rewriting
the action in terms of the bound state fields $\tilde{A}_\mu$ and $h_1$, 
giving the same result as before for the first and third terms, but for the 
kinetic energy term ${\cal A}_{KE}$, one has instead:
\begin{equation}
{\cal A}_{KE} \sim F^2 \left( \left[ \frac{g_1}{2} a_\mu + g_2 \tilde{A}_\mu
   \right]^2 \right)_{11} + (\partial_\mu h_1)^2,
\label{calAKEtilde}
\end{equation}
where the first is the mass term for the vector bound states $\tilde{A}_\mu$
which can be rewritten in the familiar form:
\begin{equation}
\sum_{\alpha,\beta} \tilde{A}_\mu^\alpha \tilde{A}^{\mu \beta} M_{\alpha \beta}
\label{Malphabeta}
\end{equation}
with $\tilde{A}^0_\mu = a_\mu$ and
\begin{equation}
M_{\alpha \beta} = \frac{F^2}{4} \left( \begin{array}{cccc}
   g_2^2 & 0 & 0& 0 \\ 0 & g_2^2 & 0 & 0 \\ 0 & 0 & g_2^2 & g_2 g_1 \\
   0 & 0 & g_1 g_2 & g_1^2 \end{array} \right).
\label{vectmass} 
\end{equation}
One recovers thus the familiar mixing between $\gamma$ and $Z$ except that
there is now a difference in the interpretation.  The mixing here is one
between the $u(1)$ gauge boson with a neutral $su(2)$ singlet bound state 
of $\Phi^\dagger \Phi$, not with the $su(2)$ gauge field.  There is thus
no breaking of the local $su(2)$ gauge symmetry.  What is broken is only
the global symmetry $\widetilde{su}(2)$, and this by the $u(1)$ interaction
which assigns to certain directions in $\widetilde{su}(2)$ a $u(1)$ charge
of $\pm g_1/2$.
This difference with the conventional formulation is already pertinent to
the confinement picture, as noted e.g.\ by 't~Hooft 
\cite{tH1}, and not special to the framon suggestion.

What is new, however, for the framon proposal, is the $\tilde{u}(1)$ invariance
which made no appearance in the standard formulation of 
electroweak theory.  Such a new
invariance would lead to a new conserved (global) charge, and it is of
interest to ask what physical significance it has, if any.  Of particles 
so far considered, only the $W^{\pm}$ carry such a charge:
\begin{eqnarray}
W_\mu^+ & = & \bphi^{\tilde{1} \dagger} (A_\mu^i \tau_i) \bphi^{\tilde{2}}
   \sim [1, -1], \\
W_\mu^- & = & \bphi^{\tilde{2} \dagger} (A_\mu^i \tau_i) \bphi^{\tilde{1}}
   \sim [- 1, 1],
\label{Wmupm}
\end{eqnarray}
with $\gamma, Z, H$ all having both $u(1)$ and $\tilde{u}(1)$ charges zero.
Since the 2 charges are always equal in magnitude and opposite in sign, 
the conservation of the one will always imply the conservation of the other
so that the $\tilde{u}(1)$ charge has so far no additional significance.

This is no longer true, however, when other particles are introduced which
do not a priori carry both charges with opposite signs.  This is the case
with fermions, which have so far been left out of our consideration but is
now of interest.  Let us introduce as usual the left-handed fermion field
$\psi_L$ as an $su(2)$ doublet.  Being of necessity a representation of the 
gauge group $U(2)$, it must then carry also a $u(1)$ charge $\pm g_1/2$, of 
which to follow the standard convention for leptons we take as $- g_1/2$.
Like the gauge fields, this $\psi$ field carries no $\tilde{u}(1)$ charge,
since this charge arises only from the introduction of the global reference
frame which affects only the framons.  Further, being an $su(2)$ doublet, 
$\psi_L$ cannot exist as a freely propagating particle in the confinement 
picture. It can, however, form bound states by $su(2)$ confinement with 
the framon field $\Phi$, thus in the ``locked gauge''
(\ref{Phiucharge})
where 
$\bphi^{(+)}$ is real and points in the up direction:
\begin{equation}
\Phi \psi_L = \left( \begin{array}{c} \nu \\ e^- \end{array} \right)
   \sim \left(\!\!\!\! \begin{array}{ccc} 
{}& [0, -\half]&{} \\ {}& [-1, \half]&{}
    \end{array} \!\!\!\!\right),
\label{chiL}
\end{equation}
where, following 't~Hooft \cite{tH,tH1}
and Banks and Rabinovici \cite{BR}, we have identified the
bound states as left-handed leptons.  In (\ref{chiL}), as before, the first 
number inside the square bracket denotes the $u(1)$ and the second the 
$\tilde{u}(1)$ charge.  We notice now that the $\tilde{u}(1)$ charge
is no longer equal in magnitude and opposite in sign to the  $u(1)$
charge as it was for the bosons, but is shifted from $-q$, the  $u(1)$
charge, by an amount $-L/2$; thus
\begin{equation}
\tilde{q} = - q - L/2,
\label{qtildeq}
\end{equation}
where we may identify $L$ as the lepton number.
Hence
the invariance of the theory under both $u(1)$ and $\tilde{u}(1)$ gives
as consequence the conservation of lepton number as well as that of the
electric charge.  This prediction of the conservation of lepton number
by the framon proposal is of no great interest when restricted to the
simple electroweak theory, since lepton number conservation there is in
any case implied by fermion number conservation.  But it will take on
a new significance when the theory is extended to include quarks in the
strong sector, leading to $B-L$ conservation,
as will be reported in a companion paper \cite{prepsm}.

\section{``Chromodynamics''}

By ``chromodynamics'' we mean here the gauge theory with gauge symmetry 
$su(3)$ which
forms the basis of the current theory of strong interactions without implying
as yet that in the way we handle it here it will be a realistic template for
such a theory.  And we shall consider it for now in isolation without
taking into account the electroweak theory to which we know it is
intimately coupled in nature.  
When considered as a theory of strong interactions, the
$su(3)$ theory is believed to be confining and as such, in the conventional
conception, does not require any scalar fields, as the electroweak
theory does, 
for symmetry breaking.  However, now that we propose that scalar fields play
in gauge theories the role of frame vectors in internal symmetry space, it
would seem more natural to introduce them in this theory as well.  Besides, 
as indicated in section 4 above, the fact that a theory is confining does 
not in itself preclude the possibility that it shares some properties of a 
broken symmetry.  Rather, if we were to introduce into the theory the framon 
scalar fields as suggested, then colour $su(3)$ confinement would bring with 
it a broken global $\widetilde{su}(3)$ symmetry, which indeed may not be so
unwelcome as a possible candidate for the three-fold symmetry of fermion 
generations, as suggested in \cite{physcons,genmixdsm}. In any case, it would 
be interesting, as we now propose to do, to consider this as a possibility. 

Let us then introduce as framon fields the elements of the matrix $\Phi$:
\begin{equation}
\Phi = (\phi_a^{\tilde{a}}),\ \ \ a =1, 2, 3; \ \ \tilde{a} = 1, 2, 3.
\label{Phisu3}
\end{equation}
Apart from requiring the determinant to be real, as explained in
section 2, 
we allow the different $\phi_a^{\tilde{a}}$ to vary independently 
over all complex values.  Next, we construct an action in accordance with 
the criterion suggested there obtaining (\ref{calA}) and (\ref{VPhi}) as 
the result.

Our first task is to examine the potential (\ref{VPhi}) to elucidate its 
properties.  For the case of $su(3)$, the
potential can be conveniently rewritten in terms of the 3-vectors in colour
space $\bphi^{\tilde{a}}$, with $\tilde{a} = 1, 2, 3$, thus:
\begin{equation}
V[\Phi] = - \mu \sum_{\tilde{a}} \bphi^{\tilde{a}}|^2
  + \lambda\,  \left(\sum_{\tilde{a}}  
|\bphi^{\tilde{a}}|^2 \right)^2
  + \kappa \sum_{\tilde{a}, \tilde{b}} 
    |(\bphi^{\tilde{a}} \cdot
       \bphi^{\tilde{b}})|^2,
\label{VPhia}
\end{equation}
And it is instructive, at least for us, to examine it in comparison with 
the potential:
\begin{equation}
V_{DSM}[\Phi] =  V[\Phi] - \kappa \sum_{\tilde{a}} 
   |\bphi^{\tilde{a}}|^4,
\label{VDSMPhi}
\end{equation}
differing from it only in that in the $\kappa$ term, the sum is taken in
$V_{DSM}$ not over all $\tilde{a}$ and $\tilde{b}$ but over only $\tilde{a}
\neq \tilde{b}$.  The latter potential $V_{DSM}$, which is not
invariant under $\widetilde{su}(3)$ as required here, was previously used 
by us in a phenomenological model
\cite{physcons,ckm,phenodsm,genmixdsm}
(which we call the Dualized Standard Model (DSM))
quite successfully, we think, to explain the fermion mass hierarchy and
mixing patterns.

First, let us summarize briefly the properties of the potential $V_{DSM}$
a slight modification of which will apply to the framon potential $V$ of 
real interest to us here.  In $V_{DSM}$, only the $\kappa$ term depends on 
the relative orientations of the vectors $\bphi^{\tilde{a}}$.
Hence, for $\kappa > 0$, the minimum of the potential occurs when these
vectors are mutually orthogonal.  In that case, the potential reduces back
to the Mexican hat potential (\ref{VMHPhi}), only now with the argument 
$|\mbox{{\boldmath $\phi$}}|$ replaced by
\begin{equation}
\zeta = \sqrt {\sum_{\tilde{a}} |\bphi^{\tilde{a}}|^2}\,.
\label{zeta}
\end{equation}
This yields as usual the minimum at
\begin{equation}
\zeta = \sqrt {\mu/(2 \lambda)},
\label{zetamin}
\end{equation}
independently of how $\zeta$ is distributed among the 3 
$|\bphi^{\tilde{a}}|$'s. In
other words, if we write $|$\bphi$^{\tilde{a}}|$ for $\tilde{a} = 1, 2, 3$ 
as a 3-vector, thus
\begin{equation} 
(|\bphi^{\tilde{1}}|, 
   |\bphi^{\tilde{2}}|,
   |\bphi^{\tilde{3}}|) =  \zeta (x, y, z), 
\label{zetaxyz}
\end{equation}
with 
\begin{equation}
x^2 + y^2 + z^2 = 1,
\label{xyznorm}
\end{equation}
the minimum of $V_{DSM}$ is degenerate with respect to the direction of the
vector $(x, y, z)$ in 3-space, or that $V_{DSM}$ has a trough or valley in
$xyz$-space with a flat bottom at a constant radius given by (\ref{zetamin}).

What happens with the framon potential in (\ref{VPhia}) of present interest?
A straightforward analysis gives that because of the term additional 
to $V_{DSM}$ as exhibited in (\ref{VDSMPhi}), the minimum will now lose its
degeneracy in the direction of the vector in (\ref{zetaxyz}) and settles
at the symmetric point $(x, y, z) = \frac{1}{\sqrt{3}}(1, 1, 1)$.  The
valley or trough found in $V_{DSM}$ is distorted by the extra term in the
framon potential so as to lose both its flat bottom and its constant radius.
Thus, if we were to start at the stationary (saddle) point at $(x, y, z) =
(1, 0, 0)$ with $\zeta = \sqrt{\mu/2(\lambda + \kappa)}$, it will roll
down the trough with changing $(x, y, z)$ and changing  radius $\zeta = 
\sqrt{\mu/(2[\lambda + \kappa (x^4 + y^4 +z^4)])}$ until it reaches the true 
minimum at $(x, y, z) = \frac{1}{\sqrt{3}}(1, 1, 1)$.

Next, we need to examine the Higgs boson spectrum.  We shall do so in the
confinement picture described in section 4.  Following the logic outlined
therein, we first fix the gauge of the local $su(3)$ symmetry by acting on 
the framon field $\Phi$ from the left by an $su(3)$ transformation rotating
$\Phi$ into a canonical form which we take to be the triangular form, 
thus:
\begin{equation}
\Phi_F = \left( \begin{array}{ccc}
   H^{\tilde{1}}_1 & \eta^{\tilde{2}}_1 & \eta^{\tilde{3}}_1 \\
   0 & H^{\tilde{2}}_2 & \eta^{\tilde{3}}_2 \\
   0 & 0 & H^{\tilde{3}}_3 \end{array} \right),
\label{Phitriang}
\end{equation}
with $H^{\tilde{a}}_a$ real and $\eta^{\tilde{a}}_b$ complex, the 8 degrees
of freedom in $su(3)$ being just sufficient to do so for $\Phi$ (in the 
``minimal'' embedding with determinant real). 

First, for the potential $V_{DSM}$, the minimum is degenerate and occurs 
in this gauge when $\Phi$ is diagonal,
\begin{equation}
\Phi_0 = \zeta \left( \begin{array}{ccc} x & 0 & 0 \\ 0 & y & 0 \\ 0 & 0 & z
              \end{array} \right).
\label{diagxyz}
\end{equation}
Hence, we can rewrite $\Phi$ as;
\begin{equation}
\Phi_F = \left( \begin{array}{ccc}
   \zeta x + h^{\tilde{1}}_1 & \eta^{\tilde{2}}_1 & \eta^{\tilde{3}}_1 \\
   0 & \zeta y + h^{\tilde{2}}_2 & \eta^{\tilde{3}}_2 \\
   0 & 0 & \zeta z + h^{\tilde{3}}_3 \end{array} \right),
\label{Phitriangp}
\end{equation}
with $h^{\tilde{a}}_a$ real and $\eta^{\tilde{a}}_b$ complex being the small 
fluctuations about the minimum, which are to represent the Higgs degrees of 
freedom. One sees thus immediately that there are to be 9 Higgs bosons in 
this scheme.

In the confinement picture, the Higgs bosons are to be considered as bound
states $\Phi^{\dagger} \Phi$ confined by $su(3)$.  To find their spectrum,
we need to evaluate $V_{DSM}$ in the triangular gauge to second order in 
the fluctuations $h^{\tilde{a}}_a$ and $\eta^{\tilde{a}}_b$.  We obtain
straightforwardly, on putting $\mu = 2 \lambda \zeta^2$:
\begin{equation}
V_{DSM}[\Phi] \sim - \lambda \zeta^4 + 4 \lambda \zeta^2 (x h^{\tilde{1}}_1
   + y h^{\tilde{2}}_2 + z h^{\tilde{3}}_3)^2 + 2 \kappa \zeta^2
     (x^2 |\eta^{\tilde{2}}_1|^2 + x^2 |\eta^{\tilde{3}}_1|^2
           + y^2 |\eta^{\tilde{3}}_2|^2).
\label{VDSMapprox}
\end{equation}
From this, we can read off the 7 massive eigenstates of the Higgs boson as 
the combination $(x h^{\tilde{1}}_1 + y h^{\tilde{2}}_2 + z h^{\tilde{3}}_3)$
plus the real and imaginary parts of $\eta^{\tilde{2}}_1, \eta^{\tilde{3}}_1$
and $\eta^{\tilde{3}}_2$.

To find the normalized eigenstates and hence the mass eigenvalues, we have
to make explicit the unitary transformation between the gauge when $\Phi$
is triangular and the gauge when the minimum remains diagonal whatever the 
fluctuations, i.e.
\begin{equation}
\Omega_{DF} \Phi_D = \Phi_F,
\label{PhiDF}
\end{equation}
where $\Phi_F$ is as given in (\ref{Phitriangp}) and $\Phi_D$ appears as:
\begin{equation}
\Phi_D = \left( \begin{array}{ccc} 
   \zeta x + \delta \phi^{\tilde{1}}_1 & \delta \phi^{\tilde{2}}_1
       & \delta \phi^{\tilde{3}}_1 \\
   \delta \phi^{\tilde{1}}_2 & \zeta y + \delta \phi^{\tilde{2}}_2
       & \delta \phi^{\tilde{3}}_2 \\
   \delta \phi^{\tilde{1}}_3 & \delta \phi^{\tilde{2}}_3
       & \zeta z + \delta \phi^{\tilde{3}}_3 \end{array} \right).
\label{PhiD}
\end{equation}
To first order in the fluctuations, $\Omega_{DF}$, which of course differs 
from the identity only to that order, reads as:
\begin{equation}
\Omega_{DF} = \left( \begin{array}{ccc}
   1 - i (\zeta x)^{-1} \delta \phi^{\tilde{1}}_{1 I} 
     & (\zeta x)^{-1} \delta \phi^{\tilde{1} *}_2
     & (\zeta x)^{-1} \delta \phi^{\tilde{1} *}_3 \\
   - (\zeta x)^{-1} \delta \phi^{\tilde{1}}_2 
     & 1 - i (\zeta y)^{-1} \delta \phi^{\tilde{2}}_{2 I}
     & (\zeta y)^{-1} \delta \phi^{\tilde{2} *}_3 \\
   - (\zeta x)^{-1} \delta \phi^{\tilde{1}}_3
     & - (\zeta y)^{-1} \delta \phi^{\tilde{2}}_3 
     & 1 - i (\zeta z)^{-1} \delta \phi^{\tilde{3}}_{3 I}
     \end{array} \right),
\label{OmegaDF}
\end{equation}
where subscripts $R$ and $I$ denote the real and imaginary parts.

From (\ref{PhiDF}), it follows then that:
\begin{eqnarray}
h^{\tilde{1}}_1 & = & \delta \phi^{\tilde{1}}_{1 R}; \ \ \ 
    h^{\tilde{2}}_2 = \delta \phi^{\tilde{2}}_{2 R}; \ \ \ 
    h^{\tilde{3}}_3 = \delta \phi^{\tilde{3}}_{3 R}; \nonumber \\ 
\eta^{\tilde{2}}_1 & = & \delta \phi^{\tilde{2}}_1 
    + \frac{y}{x} \delta \phi^{\tilde{1} *}_2; \nonumber \\
\eta^{\tilde{3}}_1 & = & \delta \phi^{\tilde{3}}_1 
    + \frac{z}{x} \delta \phi^{\tilde{1} *}_3; \nonumber \\ 
\eta^{\tilde{3}}_2 & = & \delta \phi^{\tilde{3}}_2 
    + \frac{z}{y} \delta \phi^{\tilde{2} *}_3.
\label{etaphi}
\end{eqnarray}
This gives the normalized Higgs mass eigenstates as:
\begin{eqnarray}
H_1 & = & x h^{\tilde{1}}_1 + y h^{\tilde{2}}_2 +  z h^{\tilde{3}}_3
      = x \delta \phi^{\tilde{1}}_{1 R} + y \delta \phi^{\tilde{2}}_{2 R}
          + z \delta \phi^{\tilde{3}}_{3 R} \nonumber \\
H_2 & = & \frac{y}{\sqrt{y^2 + z^2}} \eta^{\tilde{3}}_{2R}
      = \frac{1}{\sqrt{y^2 + z^2}} (y \delta \phi^{\tilde{3}}_{2R}
         + z \delta \phi^{\tilde{2}}_{3R}) \nonumber \\
H_3 & = & \frac{y}{\sqrt{y^2 + z^2}} \eta^{\tilde{3}}_{2I}
      = \frac{1}{\sqrt{y^2 + z^2}} (y \delta \phi^{\tilde{3}}_{2I}
         - z \delta \phi^{\tilde{2}}_{3I}) \nonumber \\
H_4 & = & \frac{x}{\sqrt{x^2 + z^2}} \eta^{\tilde{3}}_{1R}
      = \frac{1}{\sqrt{x^2 + z^2}} (x \delta \phi^{\tilde{3}}_{1R}
         + z \delta \phi^{\tilde{1}}_{3R}) \nonumber \\
H_5 & = & \frac{x}{\sqrt{x^2 + z^2}} \eta^{\tilde{3}}_{1I}
      = \frac{1}{\sqrt{x^2 + z^2}} (x \delta \phi^{\tilde{3}}_{1I}
         - z \delta \phi^{\tilde{1}}_{3I}) \nonumber \\
H_6 & = & \frac{x}{\sqrt{x^2 + y^2}} \eta^{\tilde{2}}_{1R}
      = \frac{1}{\sqrt{x^2 + y^2}} (x \delta \phi^{\tilde{2}}_{1R}
         + y \delta \phi^{\tilde{1}}_{2R}) \nonumber \\
H_7 & = & \frac{x}{\sqrt{x^2 + y^2}} \eta^{\tilde{2}}_{1I}
      = \frac{1}{\sqrt{x^2 + y^2}} (x \delta \phi^{\tilde{2}}_{1I}
         - y \delta \phi^{\tilde{1}}_{2I}),
\label{HK}
\end{eqnarray}
with eigenvalues read off from (\ref{VDSMapprox}) as $4 \lambda \zeta^2$
for $H_1$, $2 \kappa \zeta^2 (y^2 + z^2)$ for $H_2, H_3$, $2 \kappa \zeta^2 
(x^2 + z^2)$ for $H_4, H_5$, and $2 \kappa \zeta^2 (x^2 + y^2)$ for 
$H_6, H_7$.  Further there are 2 zero modes, say $H_K, K = 8, 9$ which are
linear combinations of $h^{\tilde{1}}_1, h^{\tilde{2}}_2, h^{\tilde{3}}_3$ 
orthogonal to $H_1$.  These results are identical, as they should be, to 
those obtained earlier in \cite{ckm} with the potential (\ref{VDSMPhi}) 
using the symmetry breaking picture. 

But what happens for the framon potential (\ref{VPhia}) of actual interest?
Following the same procedure, we first evaluate the potential for $\Phi$ 
in the canonical (triangular) gauge (\ref{Phitriang}) up to second order
in the fluctuations about the point $(x, y, z))$ as above, obtaining:
\begin{eqnarray}
V[\Phi] & \sim & - \zeta^4 [ \lambda + \kappa (x^4 + y^4 + z^4)] \nonumber \\
   & & + 4 \kappa \zeta^3 [(x^3 h^{\tilde{1}}_1 + y^3 h^{\tilde{2}}_2  
      + z^3 h^{\tilde{3}}_3) - (x^4 + y^4 + z^4)(x h^{\tilde{1}}_1 
      + y h^{\tilde{2}}_2 + z h^{\tilde{3}}_3)] \nonumber \\
   & & + 4 \lambda \zeta^2 (x h^{\tilde{1}}_1 + y h^{\tilde{2}}_2 
      + z h^{\tilde{3}}_3)^2 \nonumber \\
   & & + 2 \kappa \zeta^2 [3 (x^2 h^{\tilde{1} 2}_1 + y^2 h^{\tilde{2} 2}_2 
      + z^2 h^{\tilde{3} 2}_3) - (x^4 + y^4 + z^4)(h^{\tilde{1} 2}_1
      + h^{\tilde{2} 2}_2 + h^{\tilde{3} 2}_3)] \nonumber \\
   & & + 2 \kappa \zeta^2 [(x^2 + y^2)|\eta^{\tilde{2}}_1|^2 
     +(x^2 + z^2)|\eta^{\tilde{3}}_1|^2+(y^2 + z^2)|\eta^{\tilde{3}}_2|^2 
       \nonumber \\
   & &  - (x^4 + y^4 + z^4)(|\eta^{\tilde{2}}_1|^2 + |\eta^{\tilde{3}}_1|^2
     + |\eta^{\tilde{3}}_2|^2)]
\label{VPhiapprox}
\end{eqnarray}
We notice that $V$ in (\ref{VPhiapprox}), though more complicated, is still 
diagonal in the fluctuations $\eta^{\tilde{a}}_b$, which therefore remain 
mass eigenstates; indeed, since the transformation $\Omega_{DF}$ above and 
the subsequent arguments with it being independent of the potential, so will 
$H_K; K = 2, 3,..., 7$ remain the normalized ones.  Only in the subspace 
spanned by $h^{\tilde{1}}_1, h^{\tilde{2}}_2$ and $h^{\tilde{3}}_3$, the 
eigenstates are here no longer easy to identify, although, of course, they
remain a triad of orthonormal linear combinations of $H_1$ and the 2 zero
modes orthogonal to it.  For the purpose of the present paper, we do not 
need to specify further the actual linear combinations which occur, nor 
yet their eigenvalues, except to note that the previous zero modes would
now acquire in general nonzero masses.

The reason that we have gone to some detail in specifying the Higgs mass
eigenstates for the framon potential $V$ in comparison with those of the 
potential $V_{DSM}$, is that the latter have been instrumental in deriving 
a crucial feature in DSM \cite{physcons,ckm,
phenodsm,genmixdsm}, namely a fermion mass matrix which rotates with changing 
energy scale, which leads to a simple, yet quite successful, explanation 
for the fermion mass hierarchy and mixing patterns seen in nature.  The way 
that it works may be briefly summarized in our present language as follows.  
One starts, for reasons which need not be repeated here, with a Yukawa 
coupling term of the form:
\begin{equation}
\sum_{\tilde{a}} \bar{\mbox{\boldmath $\psi$}}_L \mbox{\boldmath $\phi$}^
   {\tilde{a}}  \sum_{[b]} Y_{[b]} \psi_R^{[b]} + {\rm h.c.},
\label{Yukawa3}
\end{equation}
where left-handed fermion field $\mbox {\boldmath $\psi$}_L$ is an $su(3)$ 
triplet and the right-handed fields  $\psi_R^{[b]}$ are $su(3)$ singlets, 
coupled to the framon fields $\bphi^{\tilde{a}}$ via the 
Yukawa couplings $Y_{[b]}$.  This gives at tree level a factorized 
fermion mass 
matrix, which by a relabelling of the right-handed fermion fields, can be 
written in a hermitian form without $\gamma_5$ as: 
\begin{equation}
m \sim m_T \left( \begin{array}{c} x \\ y \\ z \end{array} \right) (x,y,z),
\label{fmassmat}
\end{equation}
with $(x, y, z)$ having the same meaning as in (\ref{zetaxyz}).  This mass 
matrix, being of rank one, has only one nonzero eigenvalue with
eigenvector $(x, y, z)$.  
Besides, 
this eigenvector 
being the same for all fermion species, i.e.\ whether up or down 
type quarks, or charged leptons or neutrinos, the CKM and MNS mixing matrices 
are both the identity, and is thus not bad as a zeroth order approximation 
to reality, given the observed hierarchical masses and small mixing angles, 
the latter at least for quarks.  When one turns on the loop corrections,
however, the matrix $m$ remains factorized, but the factor $(x, y, z)$ now 
changes its orientation (rotates) with changing scale.  This causes the 
mass in the heavy generation to ``leak'' into the 2 lower generations 
giving them small but nonzero masses, hence the observed mass hierarchy.  
At the same time, the state vectors of the up and down states lose their 
mutual alignment at tree level and acquire thereby a nontrivial mixing 
matrix between them.  Furthermore, the rotation is found to have a fixed 
point at infinite energy at $(x, y, z) = (1, 0, 0)$ and another at zero 
energy at $(x, y, z) = \frac{1}{\sqrt{3}}(1, 1, 1)$, which offers an 
immediate qualitative explanation for some well-known but at first 
sight puzzling patterns in the measured mass and mixing parameters, 
such as that $m_c/m_t < m_s/m_b < m_\mu/m_\tau$, or that in the CKM matrix,
the elements $V_{td}, V_{ub}$ are much smaller than the others, while in
the MNS matrix, the CHOOZ angle $\theta_{13}$ is near zero but the SuperK
angle $\theta_{23}$ is near maximal.  Indeed, with only 3 real parameters
fitted to $m_c/m_t, m_\mu/m_\tau$ and the Cabbibo angle, one is able to
give a reasonable, often near quantitative, description of the fermion mass 
hierarchy, quark mixing and neutrino oscillation. 

The crucial feature of mass matrix rotation in DSM
arises from insertions into the fermion propagator of loops of the Higgs 
states listed in (\ref{HK}).  The couplings of these to the fermions can
be obtained from the Yukawa coupling term given in (\ref{Yukawa3}) and are
found also to have the factorized form:
\begin{equation}
{\Gamma}_K = |v_K \rangle \langle v_0| \frac{1}{2} (1 + {\gamma}_5)
           +  |v_0 \rangle \langle v_K| \frac{1}{2} (1 - {\gamma}_5),
\label{GammaK}
\end{equation}
where 
\begin{equation}
|v_0 \rangle = \left( \begin{array}{c} x \\ y \\ z \end{array} \right);
\ \ \ 
|v_K \rangle = \left( \begin{array}{c} 
   \sum_{\tilde{a}} (V_K)_1^{\tilde{a}} \\ 
   \sum_{\tilde{a}} (V_K)_2^{\tilde{a}} \\ 
   \sum_{\tilde{a}} (V_K)_3^{\tilde{a}}
   \end{array} \right),
\label{vangle}
\end{equation}
with $(V_K)_a^{\tilde{a}}$ being the coefficient of $\delta \phi_a^
{\tilde{a}}$ in the expressions (\ref{HK}) of the Higgs states $H_K$ in
terms of the latter variables.  With the couplings $\Gamma_K$, it is
straightforward to calculate the one-Higgs-loop corrections to the fermion
mass matrix and then to extract the scale-dependent terms of the vector
${\bf r} = (x, y, y)$.  One obtains then for ${\bf r}$ the RG equation:
\begin{equation}
\frac{d\, {\bf r}}{d(\ln \mu^2)} \sim -\sum_K \langle v_0|v_K \rangle 
   |v_K \rangle,
\label{RGE}
\end{equation}
which governs the scale dependence of the rotating vector ${\bf r}$. One
sees that there will be rotation so long as the ``governing vector'' on 
the right is neither zero nor parallel to the vector $|v_0 \rangle$. 
Thus, for example, it turns out that the ``governing vector'' vanishes at 
${\bf r} = \frac{1}{\sqrt{3}}(1, 1, 1)$, but equals $(1, 0, 0)$ when 
${\bf r}$ itself is at this value, hence the fixed points noted above.

So much then for the old phenomenological model, but what happens with
the framon potential of present interest?  We note that the so-called 
``governing vector'' for the rotation does not depend on the Higgs masses,
only on the Higgs states listed in (\ref{HK}).  But as already observed
before, the state vectors for $H_K; K = 2, ..., 7$ remain unchanged for
the framon potential, which are the only ones to give rise to rotation.
The others labelled as $H_K; K = 1, 8, 9$ are changed but remain just a triad 
of orthonormal linear combinations of the states $h_a^{\tilde{a}},\ a =
1, 2, 3$. These last all had the matrices $V_K$ appearing in (\ref{vangle})
real and diagonal, which means in turn that the vectors $|v_K \rangle$
for $K = 1, 8, 9$ themselves form a real orthonormal triad, from which it 
then follows that, whether in the framon potential or in $V_{DSM}$, 
they will give a contribution to the ``governing vector'' proportional to 
$|v_0 \rangle$ and can give no rotation.  We conclude therefore that
despite the differences, the rotational properties obtained before for 
$V_{DSM}$ still apply to the framon potential, and hence also the good
phenomenological results, if interpreted in the same way.    

It may
thus seem that if in the $su(3)$ theory one adopts the same framon 
idea as in the previous cases considered, one would end up with a
scheme with similar phenomenological achievements as the earlier model
DSM constructed specifically for the purpose.  This conclusion would
be incorrect, 
however, 
for several very strong reasons, and one cannot regard this framon scheme 
yet as anywhere near a realistic description of the stated phenomena.  
First, in our analysis of the Higgs spectrum for the framon potential 
$V[\Phi]$,  
we have expanded the action about a general point $\zeta(x, y, z)$ in the 
trough of $V[\Phi]$, but this is not a minimum of the potential as it was 
for $V_{DSM}$, of which fact we are reminded of course by the 
nonvanishing term in (\ref{VPhiapprox}) linear in the fluctuations,
and we 
have given as yet no justification why one could do so.  Secondly, the 
Yukawa coupling term (\ref{Yukawa3}) that we have copied from DSM is only
permutation symmetric in the index $\tilde{a}$, not invariant under the
global symmetry $\widetilde{su}(3)$ that the framon framework would want.
Indeed, given the quantities so far introduced in the scheme, there seems
no possibility to construct such an invariant Yukawa term, there being
no other
vector in $\widetilde{su}(3)$ space to saturate the $\tilde{a}$ indices
occurring in $\mbox{\boldmath $\phi$}^{\tilde{a}}$.  Thirdly, if we were 
to accept the confinement picture of symmetry breaking that is here adopted, 
the fermions described above are to be considered as compound states of 
the fundamental fermion fields with the scalar framon fields, i.e.\  
$\bar{\mbox{\boldmath $\psi$}}_L \Phi$, confined by $su(3)$ colour, and 
hence should be interpreted as hadrons, not the near point-like leptons 
and quarks in which we are interested.  

Nevertheless, the similarity of the above framon scheme in the $su(3)$ 
theory to DSM is indicative and offers us hope 
that when other relevant features which have not yet been included are 
taken into account, then a more realistic scheme will emerge.  Indeed, 
in a companion paper \cite{prepsm}, it is shown that when applied to
chromodynamics, not in isolation but coupled to electroweak theory
as it is in nature, then the framon idea gives
a much more realistic model with all the 
above 3 shortcomings removed.  
Still the treatment here of ``chromodynamics''
in isolation is instructive as a dry-run, first to test the waters, and
secondly to lay the ground work for an attack on the problem in the 
realistic case.

\section{Remarks}

In summary, it would seem that the idea of assigning to Higgs fields the 
geometrical significance of frame vectors in internal symmetry space in 
the manner suggested is not in contradiction with the use to which Higgs 
fields are commonly put in particle physics.  

On the practical side, it is seen that in the special case of the $u(1) 
\times su(2)$ theory, the framon idea has led uniquely, modulo only a 
hypothesis on ``minimality'' in representation and embedding, to the 
standard electroweak theory which is the only application so far of the 
Higgs mechanism to particle physics with definitive success.  When applied 
to chromodynamics, the introduction of the framon fields, which we stress
is in itself not in any contradiction to colour being confined, leads
automatically to the existence of a global 3-fold symmetry which
can play the role of fermion generations.  Although the simplified framework 
so far examined which takes account only of chromodynamics in isolation is 
not realistic, it already shows intriguing similarities to an earlier 
phenomenological model which has had good success in explaining the main 
features of the fermion generation puzzle, including in particular the 
fermion mass hierarchy, quark mixing and neutrino oscillations.  An 
attempt to construct with the framon idea a new realistic framework for 
fermion generations has met with some success and 
is reported on separately \cite{prepsm}.
  
On the theoretical side, on the other hand, the above considerations lead to 
some quite revolutionary possibilities.  At present, the standard attitude 
towards Higgs fields is that they are a tool for symmetry breaking.  We turn 
to experiment or rely on some other justifications to decide which of the 
gauge symmetries that we have started with are to be broken and in what way, 
and then we introduce into our theory the appropriate Higgs fields 
to implement 
the required breaking mechanism.  However, if we accept the outlook adopted 
in the discussion above, it would appear that the scalar fields appearing 
in a gauge theory have a geometric function of their own to discharge and 
hence are not to be introduced or discarded at will to suit our interpretation 
of experiment or some other considerations.  

Further, even the concept of whether a gauge symmetry is broken has to be 
reassessed.  As the example in section 3 above has shown, the conventional 
picture of a local gauge symmetry being spontaneously broken by the 
introduction of Higgs fields can be given an entirely different, but in 
actual application completely equivalent, interpretation, namely that the 
local gauge symmetry is confining, only the global symmetry, necessarily 
associated with it by virtue of the frame vector interpretation of Higgs 
fields, can be broken by the process and gives rise to the symmetry breaking 
phenomenon observed.  In other words, if this alternative interpretation 
of symmetry breaking is adopted, whether a gauge symmetry in a theory is to 
be broken or not is not up to us to impose but is a matter to be decided by 
the internal consistency of the theory via the interaction between the gauge 
vector and ``framon'' scalar fields.  And depending on the structure
of the gauge symmetry the physical consequences can be vastly
different, as we see in the case of $su(2)$ and $su(3)$.        

Finally, pushing the idea to the extreme limit, one could consider the 
``framon'' fields to be part and parcel of a gauge theory without which 
it may be structurally incomplete.  In that case, even the 
presence of the ``framon'' fields would be inherent in the theory, which 
in turn would decide via its own dynamics which symmetries if any are to 
be ``broken'', leaving nothing but the choice of the starting gauge symmetry 
to be injected from experiment.  We do not know whether this last extreme 
view can be maintained.  But in the case of gravity, one has certainly 
introduced the vierbeins, which are the equivalents of the ``framon'' 
scalars here, as independent variables in addition to the spin connections, 
which are the equivalents of the gauge vector fields.  And in the few 
examples of gauge theories we have chosen to study in this paper, we have 
not yet come across a blatant contradiction to this extreme viewpoint.

\end{document}